\documentstyle[11pt,IAUS215,twoside,epsfig]{article}

\def\edcomment#1{\iffalse\marginpar{\raggedright\sl#1\/}\else\relax\fi}
\reversemarginpar

\begin{document}
\vspace*{1cm}
\title{The Effects of Stellar Yields with Rotation on Chemical Evolution Models}
 \author{Leticia Carigi}
\affil{Instituto de Astronom\'{\i}a UNAM, A.P. 70-264, DF 04510, Mexico}
\author{Max Pettini}
\affil{Institute of Astronomy, Madingley Road, Cambridge CB3 0HA, England}

\begin{abstract}

We summarise the results of recent work investigating the effects
of stellar yields which take into account rotation on the 
predictions of chemical evolution models for early populations
of galaxies.

\end{abstract}

As we have heard at this meeting, stellar rotation
can affect significantly the chemical yields of stars over a wide
range of stellar masses, from the most massive stars which
drive powerful winds to the intermediate and low mass stars
which evolve to the planetary nebula stage.
We have begun to investigate some of these effects by
constructing chemical evolution models for galaxies 
using rotation- and metallicity-dependent yields and 
comparing their predictions
with those of models which do not include rotation in the 
adopted yields.

In particular we follow the detailed 
evolutions of the C/O vs O/H and N/O vs O/H ratios
and compare them with available
data for two populations of galaxies observed at
high redshift ($z \simeq 3$), when the universe (and presumably the 
galaxies) were young. The two populations are
the Damped Ly$\alpha$ Systems (DLAs)
seen in absorption in the spectra of bright
background QSOs, and 
Lyman Break Galaxies (LBGs), star-forming galaxies
recognised directly via their integrated stellar light.

Full description of the models is given in a forthcoming 
paper; here we summarise their principal features.
LBGs and DLAs are treated as closed systems (i.e. no infall nor 
outflows are included in the models). In both cases we
assume a total baryonic mass of $10^{10}\,M_{\sun}$
with initial primordial composition.
Their star formation rates (SFR) are proportional to gas mass
(${\rm SFR} = \nu M_{\rm gas}$), where the efficiency $\nu$ is constant 
in time and depends on the type of galaxy. For
LBGs we have assumed a high star formation efficiency
with $\nu = 1.0\,{\rm Gyr}^{-1}$, while $\nu$ is ten times lower in DLAs.
Qualitatively, this distinction is consistent with 
the lower luminosities and metallicities of DLAs,
compared with LBGs at the same cosmic epochs (e.g. Pettini 2001).

We adopted the stellar yields by Meynet \& Maeder (2002) 
for stars of masses between 2 to 60\,$M_{\sun}$ and for three values 
of metallicity, $Z = 10^{-5}$, 0.004, and 0.02 (respectively 
1/2000, 1/5 and 1 times solar metallicity). We interpolated 
linearly in metallicity and mass between the values published by 
Meynet \& Maeder (2002).
The main results of our models are illustrated in Figure 1,
and can be summarised as follows.

\noindent 1. Models with yields which take into account stellar
rotation are successful in reproducing the N/O ratios
measured in DLAs (and local H II regions),
whereas models without rotation do not match the data.

\noindent 2. Our best fit to the DLA observations is obtained by adopting
the yields with rotation and the initial mass function (IMF) 
by Kroupa, Tout \& Gilmore (1993).

\noindent 3. This same combination of yields and IMF can also reproduce the
typical C/O ratio recently estimated for LBGs. Although the C/O ratio is
apparently less sensitive to the effects of stellar rotation on
the yields, models with rotation still seem to provide a
better fit to the observations than those without.
%
%
\begin{figure}[]
\vspace*{0.5cm}
\centering
\resizebox{0.45\textwidth}{!}{\includegraphics{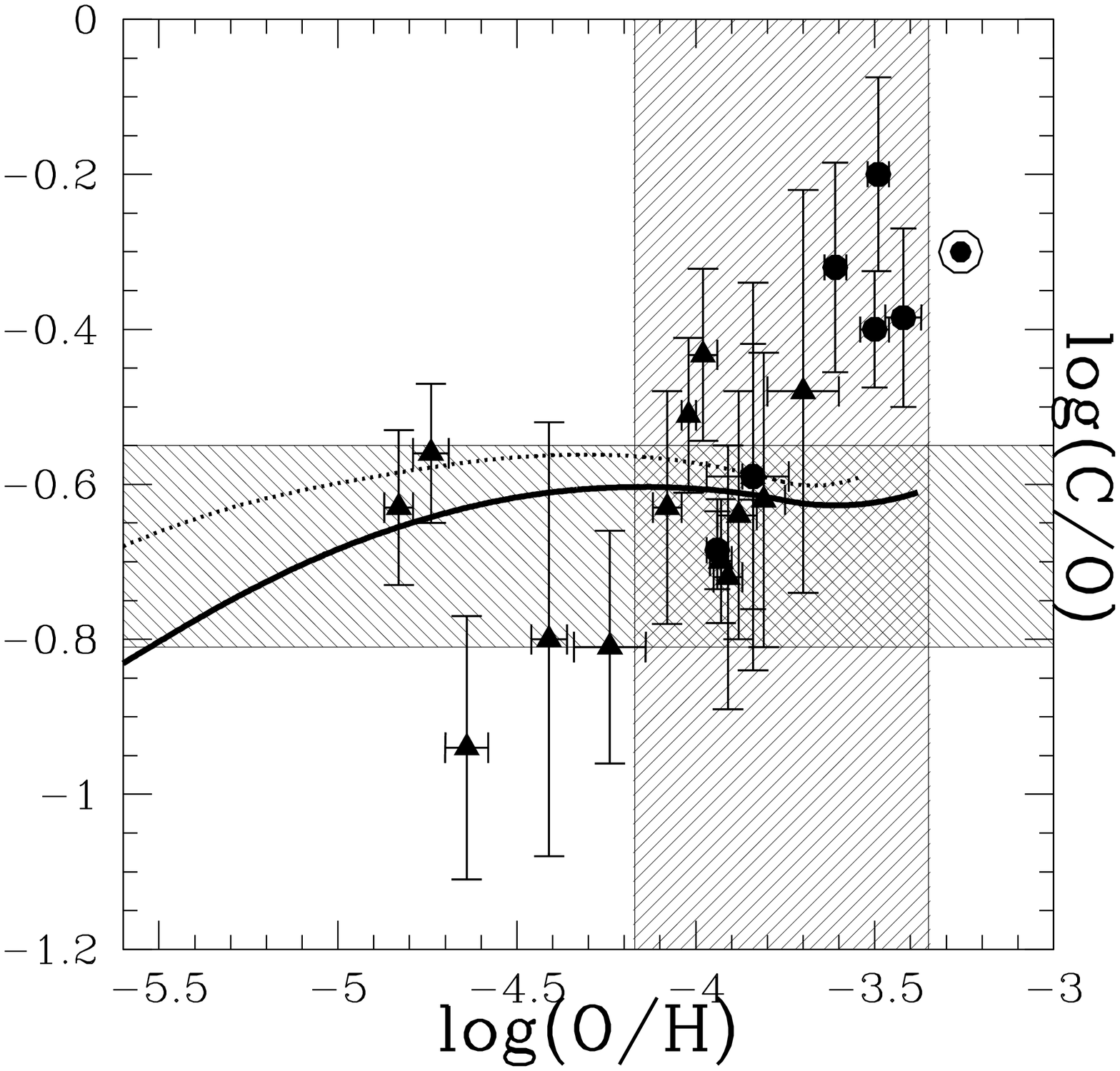}}
\resizebox{0.45\textwidth}{!}{\includegraphics{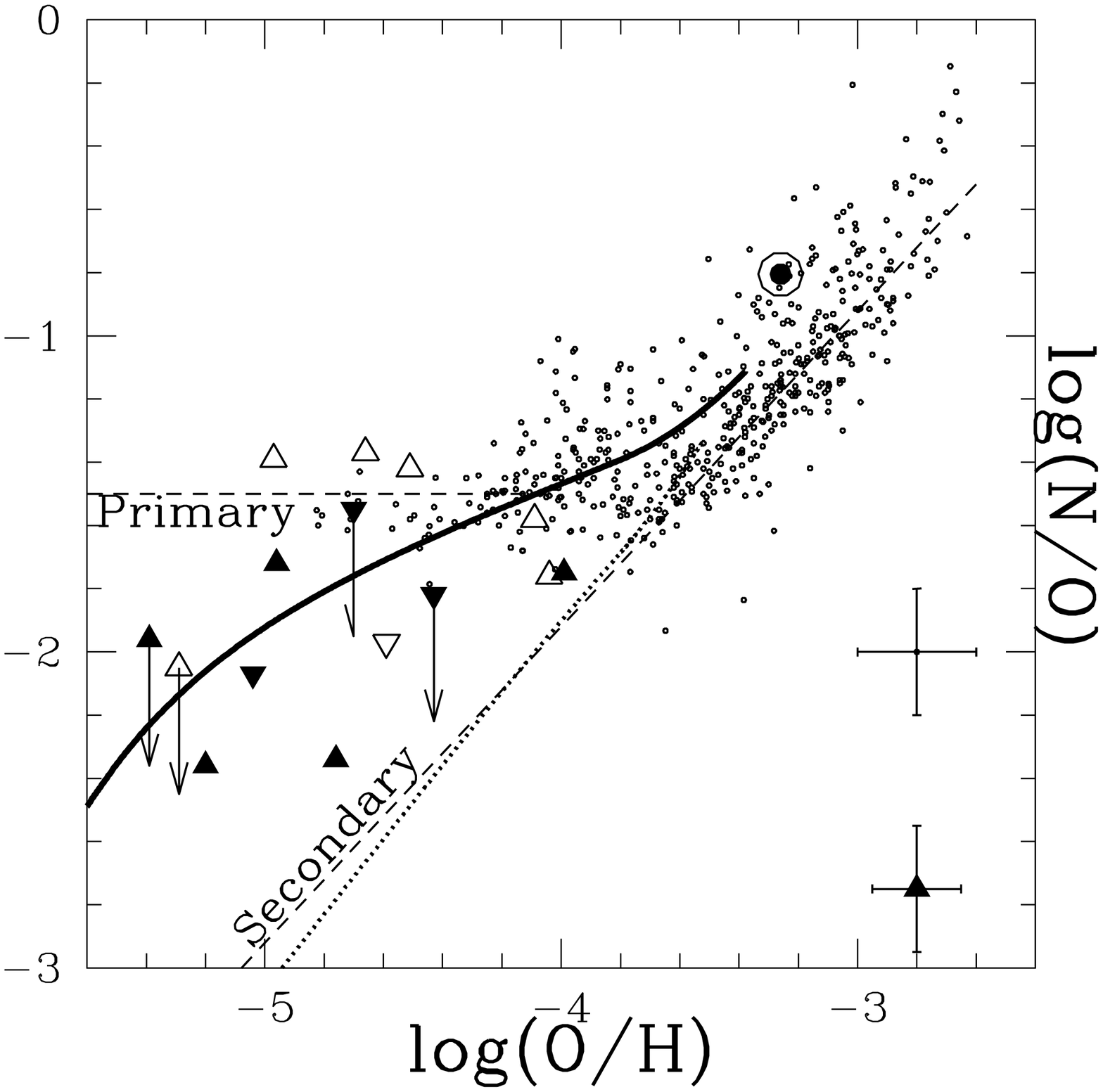}}
\caption{
{\it Left:\/} 
$\log({\rm C/O})$ vs. $\log({\rm O/H})$ for LBGs and local HII regions
(reproduced from Shapley et al. 2003, where full
references to the data are given).
Triangles are data from nearby dwarf Irregular galaxies
and the Magellanic Clouds.
Spiral galaxy data are shown with circles.
The shaded areas indicate the likely ranges of values 
of  $\log({\rm C/O})$ and $\log({\rm O/H})$ 
in LBGs.
{\it Right:\/} 
Abundances of N and O in extragalactic H~II regions
(small dots) and damped Ly$\alpha$ systems (large triangles),
reproduced from Pettini et al. (2002) (where full references
to the data can be found).
The dashed lines are approximate
representations of the secondary and primary
levels of N production.
In both plots the solid and dotted curves show
the predictions of our chemical evolution models using 
yields with and without rotation respectively.
Solar abundances are indicated by the large bulls-eye. 
}
\end{figure}

\end{document}